# Towards the optimal energy of the proton driver for a neutrino factory and muon collider

J. Strait, N. V. Mokhov, and S. I. Striganov
*Fermi National Accelerator Laboratory, P.O. Box 500, Batavia, Illinois 60510, USA*

Cross section data from the HARP experiment for pion production by protons from a tantalum target have been convoluted with the acceptance of the front-end channel for the proposed neutrino factory or muon collider and integrated over the full phase space measured by HARP, to determine the beam-energy dependence of the muon yield. This permits a determination of the optimal beam energy for the proton driver for these projects. The cross section data are corrected for the beam-energy dependent amplification due to the development of hadronic showers in a thick target. The conclusion is that, for constant beam power, the yield is maximum for a beam energy of about 7 GeV, but it is within 10% of this maximum for $4 < T_{beam} < 11$ GeV, and within 20% of the maximum for $T_{beam}$ as low as 2 GeV. This result is insensitive to which of the two HARP groups' results are used, and to which pion generator is used to compute the thick target effects.

## I. INTRODUCTION

One of the important design parameters of a possible future neutrino factory (NF) or muon collider (MC) is the energy of the high-power proton accelerator that will be used to produce the pions, whose decay muons will be captured, cooled, and stored in a storage ring, either to produce intense neutrino beams or to provide $\mu^+\mu^-$ collisions. Until recently, the study of the yield of captured muons as a function of proton beam energy has had to rely on simulations [1,2], since data on pion production cross sections over the relevant phase space has been quite sparse. Recent publication of data from the large angle spectrometer of the HARP experiment [3,4] makes it possible to address this question with experimental data.

In this paper we first present a calculation of the acceptance of the NF/MC front-end channel using the MARS15 code [5]. The acceptance is defined to be the number of muons (or pions), as a fraction of the number of pions produced at the target, that reach the end of the 50-m long tapered solenoid channel. It is computed in terms of the momentum and angle, $p$ and $\theta$, of the pions leaving the target. The acceptance is then convoluted with the measured double-differential cross section of pion production from a tantalum target, which is close in atomic weight to mercury, the favored target material for the NF/MC projects [1]. The acceptance-weighted cross section is integrated over the measured phase space, and divided by the beam kinetic energy, to give a value proportional to the muon yield normalized to constant proton beam power. Finally, corrections are made for the phase space not covered by the HARP results, and for the effects of hadronic showers that develop in a thick target, and which are not accounted for in the pure cross section data. The beam-power normalized muon yield is presented as a function of incident proton kinetic energy between 2.2 and 11.1 GeV ($3 \leq p \leq 12$ GeV/c), which brackets the beam energies under consideration for high-power proton sources at Fermilab [6] and CERN [7].

## II. FRONT-END CHANNEL ACCEPTANCE

The function of the NF/MC front-end channel is to efficiently capture pions exiting the target in the forward hemisphere, to provide a channel in which the pions decay to muons, and to allow adequate distance for a correlation to develop between the energy of the particles and their time of arrival at the end of the channel. The front-end channel model [1] used in this calculation is shown in Fig. 1.

It consists of a series of solenoids, represented by the blocks, which generate a magnetic field of 20 T at the center of the pion production target, and which drops smoothly to 1.25 T for $z > 20$ m, as shown in Fig. 2. The solenoid channel is filled with a heavy absorber, which has an inner radius of 7.5 cm around the target, and which grows linearly to 30 cm for $z > 18.62$ m. In this paper, the material within the solenoid channel is treated as a perfect absorber. It was found in earlier studies [1] that the contribution of pions scattered from the innermost material to the yield at the end of the channel was at the level of a few percent, since secondary particles that reenter the channel are efficiently swept back into the absorber by the solenoidal field. Therefore, this simplifying assumption has practically no effect on the computed acceptance.

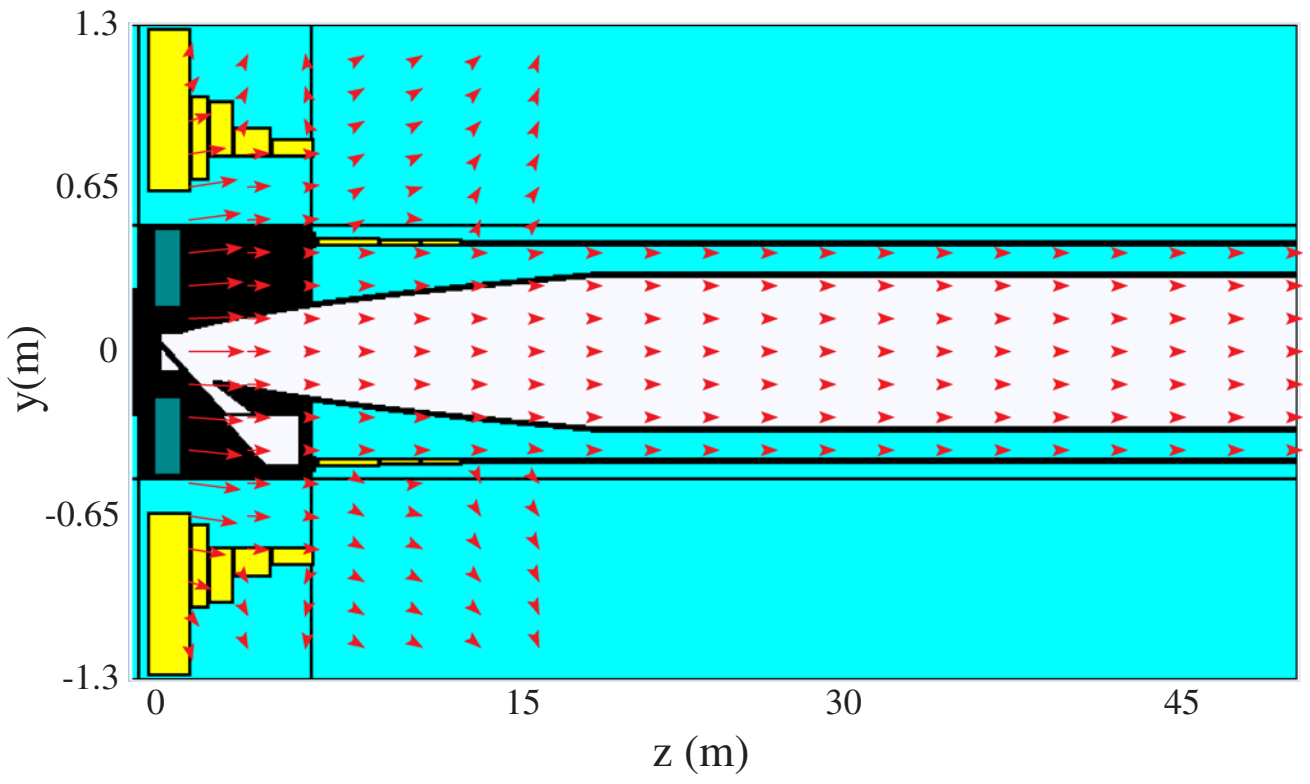

FIG. 1. MARS model of the NF/MC front-end channel. The yellow blocks represent the superconducting solenoid coils while the dark-blue ones are for the normal conducting coils of the 20-T hybrid solenoid. The red arrows indicate the strength and direction on the magnetic field. The black area represents the mercury jet target system with its absorber.

It should be noted that this channel was optimized for the proton beam energies of 16 GeV [8] and 24 GeV [9]. Although, the solenoid channel parameters optimized for the capture of the muons of interest (see below) are not very sensitive to the proton beam energy [10], the front-end channel could need to be fine-tuned for the lower energies considered in this paper.

In this paper, to compute with MARS15 the acceptance, only the solenoid channel is included; the mercury jet target is not simulated. Pions are generated from a point source at a position where the beam would exit the downstream end of the target, 27 cm downstream of the center of the first, high-field solenoid that surrounds the target. The pions are tracked through the solenoid channel and allowed to decay. The decay muons are tracked to the end of the channel, 50 m downstream of the target. The acceptance is defined as the probability that a pion of a given angle and momentum at the target yields a surviving muon (or pion) at the end of the channel. In addition, to be accepted, the muon (or pion) must have a kinetic energy in the range $40 < T_\mu < 180$ MeV, in order to be efficiently captured by the downstream rf channel [10]. The acceptance function is expressed in terms of the kinematic variables of the pion as it exits the target, in the coordinate system of the solenoid channel. Because this is a straight, axially symmetric solenoid channel, the acceptance is identical for both pion signs.

The acceptance, $A$, is shown in Fig. 3 as a function of pion momentum and angle at the target. The acceptance is large ($> 70\%$) for $150 < p_\pi < 300$ MeV/$c$ and $\theta_\pi < 0.5$ rad, and for $\theta_\pi < 1.1$ rad and $150 < p_\pi < 200$ MeV/$c$. The falloff in acceptance at low momenta results from the requirement that the muon kinetic energy is at least 40 MeV (100 MeV/$c$). The drop in acceptance at high momenta comes primarily from the requirement that $T_\mu < 180$ MeV (265 MeV/$c$), and secondarily from the transverse momentum $p_T < 225$ MeV/$c$ that is captured by the target solenoid ($B_z = 20$ T, $r = 7.5$ cm). The acceptance extends a bit beyond $\theta = \pi/2$ since the pions are generated slightly downstream of the center of the target solenoid, where the magnetic field is maximum, resulting in a small "magnetic mirror" effect. The natural edges of the acceptance are smeared by the kinematics of the $\pi \to \mu\nu$ decay.

Figure 3 also shows the kinematic regions covered by the analyses performed by the main HARP collaboration [3] and the HARP-CDP group [4]. The HARP-CDP group presents its results in terms of $p_T$ rather than $p$, resulting in the nonlinear boundary in Fig. 3. The lower limit in $p_T$ and $\theta$ of the published HARP-CDP results varies somewhat from data set to data set. The region above and to the right of the boundary represents the largest region that is covered by their results for all beam momenta $3 \leq p_{\rm beam} \leq 12$ GeV/c on a tantalum target and for both $\pi^+$ and $\pi^-$ final state particles.

Weighted by the differential phase space $2\pi \sin\theta d\theta dp$, the region analyzed by HARP (HARP-CDP) covers 87% (65%) of the front-end channel acceptance. Thus the measured pion production cross sections, weighted by the acceptance, can give a good estimation of the beam-energy dependence of the muon yield even with no corrections for the fact that these data do not cover the forward region, $\theta < 350$ mrad.

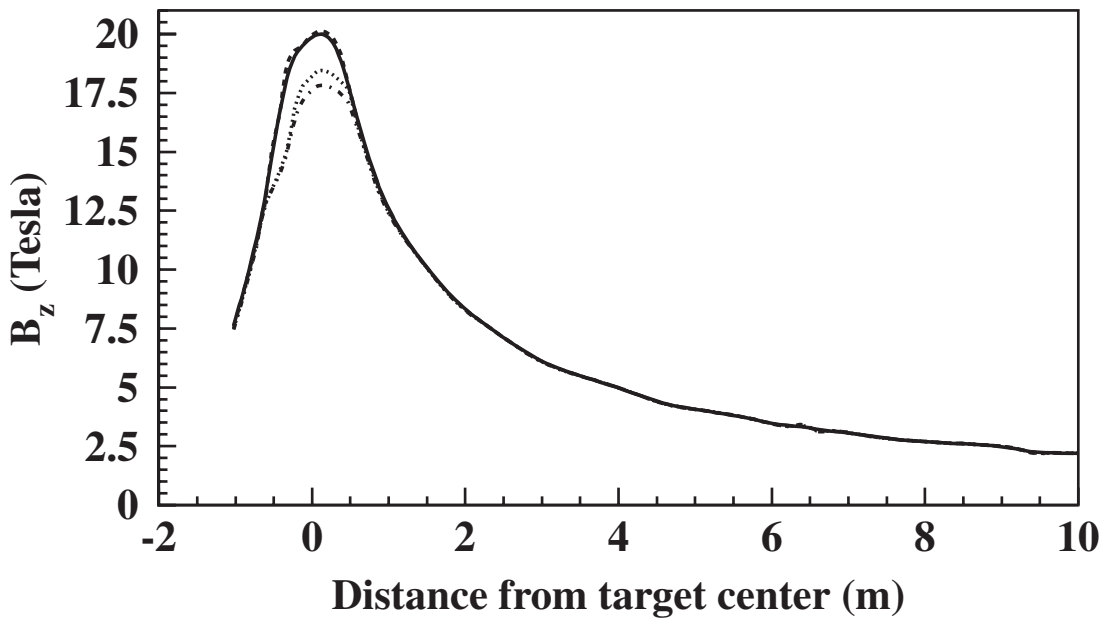

FIG. 2. Longitudinal magnetic field in the NF/MC front-end channel. The lines represent various radial positions $r$(cm) = 0 (solid), 15 (dashed), 21 (dotted), and 27 (dot-dashed).

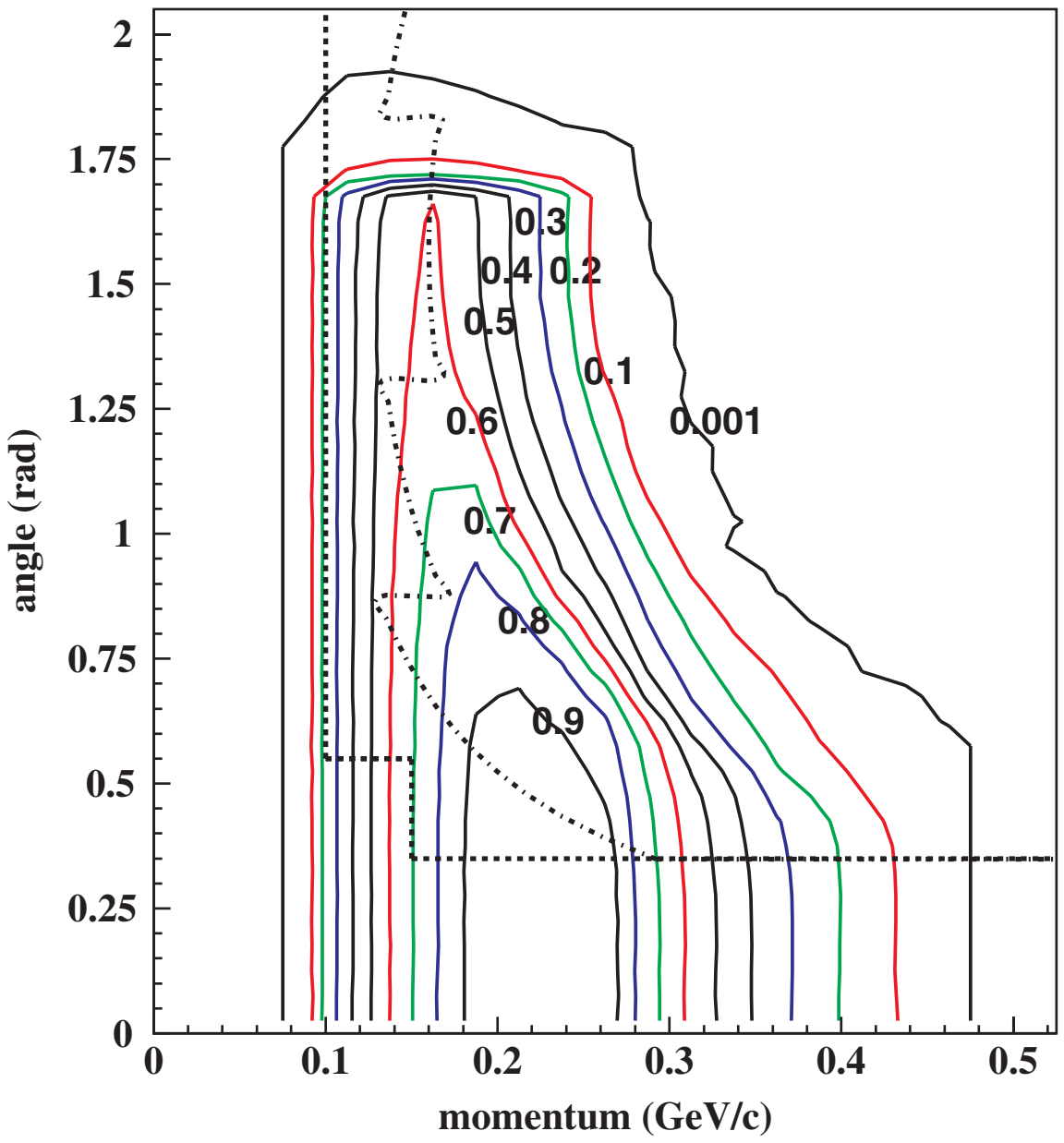

FIG. 3. Acceptance, $A$, of the NF/MC front-end channel, expressed in terms of the kinematic variables of the pion as it exits the production target. The kinematic region analyzed by the HARP (HARP-CDP) collaboration is the region above and to the right of the thick dotted (solid) line.

## III. ENERGY DEPENDENCE OF THE INTEGRATED CROSS-SECTIONS

The favored target material for the neutrino factory and muon collider is liquid mercury [1], although solid targets made of other heavy nuclei, such as tantalum or tungsten, are still under consideration [2]. For this study we have used the cross sections measured by HARP for tantalum. This is the heaviest target for which published results are available from both HARP groups, allowing an important check on the sensitivity of the conclusions to which analysis of the HARP raw data is used. The double-differential cross sections are weighted by the acceptance of the front-end channel, and integrated over the full published data set. The integrated, acceptance-weighted cross sections are then divided by the beam proton kinetic energy to give a measure of the muon yield at constant beam power.

The propagation of the errors, from the individual points to the acceptance-weighted integrated cross section, must take account of the significant systematic errors on each point. These are typically larger than the statistical errors for individual differential cross section measurements, and are substantially correlated among the individual points. In practice, this means that the error propagation can only be accurately done by the HARP groups themselves. Both the HARP [11] and HARP-CDP [12] groups have done this calculation for us.

The acceptance-weighted cross sections, divided by beam energy, for $\pi^+$ and $\pi^-$ production by protons off of tantalum are shown in Figs. 4 and 5 as a function of the beam energy for the HARP and HARP-CDP results, respectively. Three error bars are shown for each point: the innermost is the statistical error; the middle includes the systematic errors that do not correlate among the individual points, added in quadrature with the statistical errors; the outermost includes the statistical error and all systematic errors, including those that are fully correlated among the data points. Where only one or two error bars are visible, this is because the smaller errors are smaller than the data point.

The beam power normalized, acceptance-weighted cross section is relatively insensitive to the beam energy between 2 and 11 GeV. The $\pi^-$ yield is largest at 4.1 GeV (5 GeV/c), and is within 10% of the maximum at 2.2 and 7.1 GeV (3 and 8 GeV/c). For $\pi^+$ production, the maximum yield is at 2.2 GeV, although the yield, based on the HARP results, is essentially the same at 4.1 GeV. The yield is down by 12% at 7.1 GeV, based on the HARP results, and 20% based on the HARP-CDP analysis.

The overall acceptance-weighted cross section is 30%–40% lower using the HARP-CDP data than using the HARP results. This is principally a consequence of the smaller phase space analyzed by the CDP group (see Fig. 3). When the HARP cross sections are integrated only over the region covered by the CDP analysis, the $\pi^-$ results from the two groups are very similar: the CDP results are 5%–6% higher for 2–7 GeV and 15% higher for 11 GeV. For $\pi^+$, the difference is larger: the CDP cross section is lower by about 20% for the three higher energy points and 13% lower at the lowest energy. This is qualitatively consistent with the difference between the two groups regarding particle identification, in which HARP identifies some particles as pions, which CDP says are protons.

An overall figure of merit for the neutrino factory is the sum of $\pi^+$ and $\pi^-$ cross sections: $Y_P = (A\sigma^+ +$

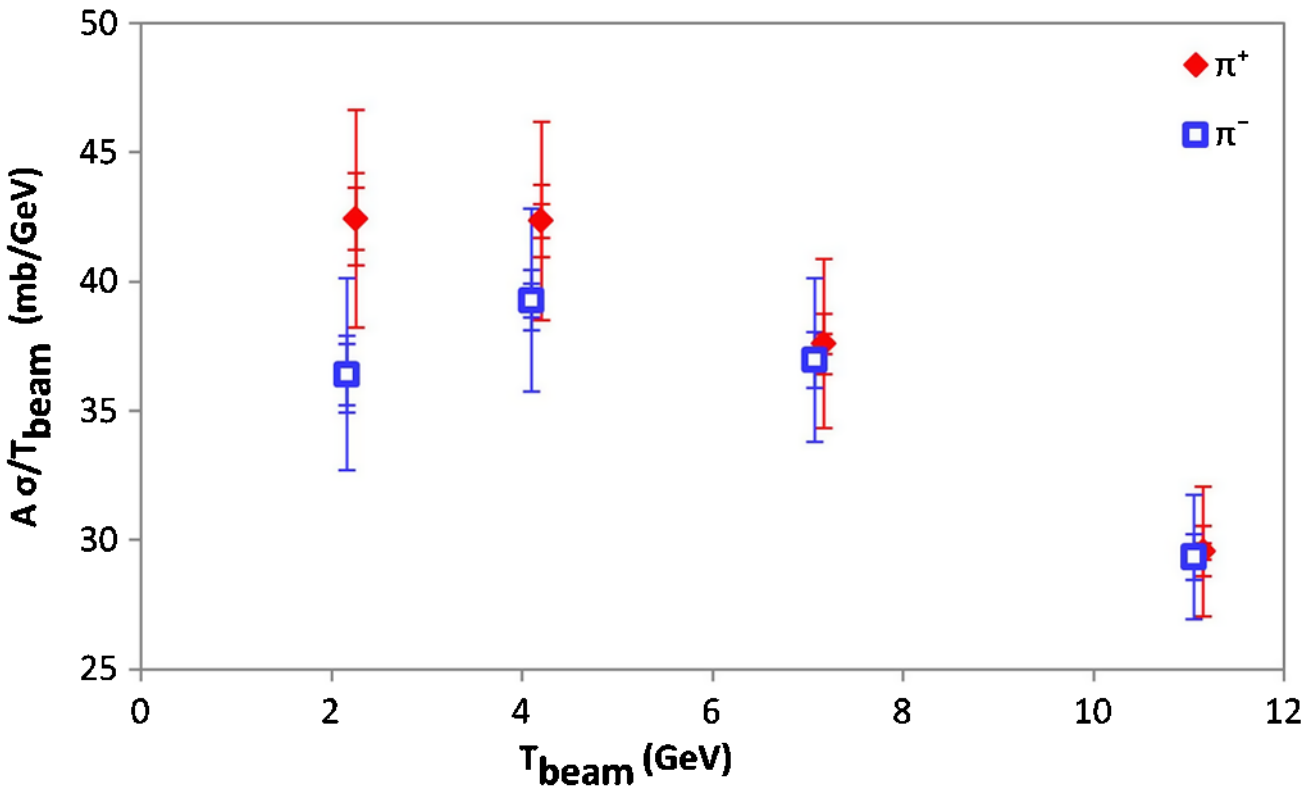

FIG. 4. Acceptance-weighted cross sections, divided by beam energy, for $\pi^+$ and $\pi^-$ production by protons off of tantalum, using data from the HARP group. The points for $\pi^+$ and $\pi^-$ have been displaced, respectively, left and right by 50 MeV for clarity of presentation.

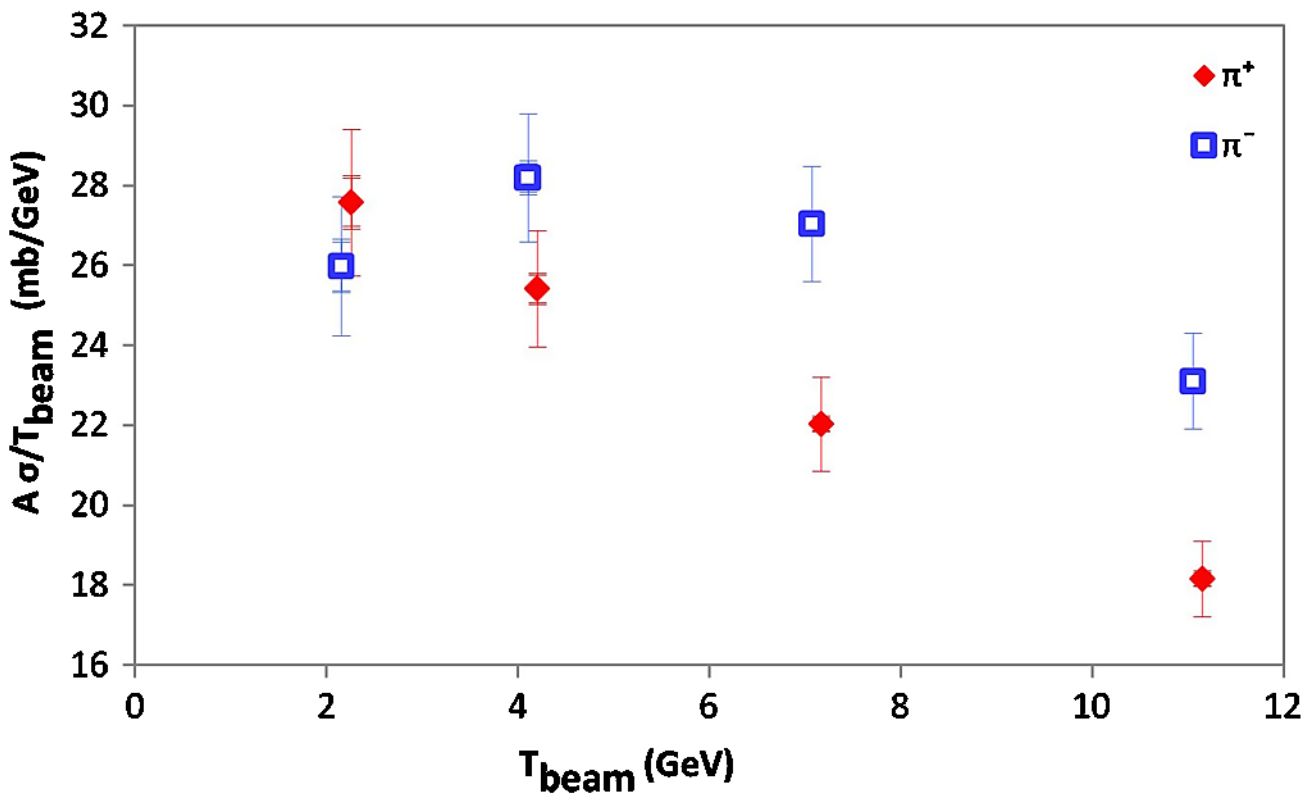

FIG. 5. Acceptance-weighted cross sections, divided by beam energy, for $\pi^+$ and $\pi^-$ production by protons off of tantalum, using data from the HARP-CDP group. The points for $\pi^+$ and $\pi^-$ have been displaced, respectively, left and right by 50 MeV for clarity of presentation.

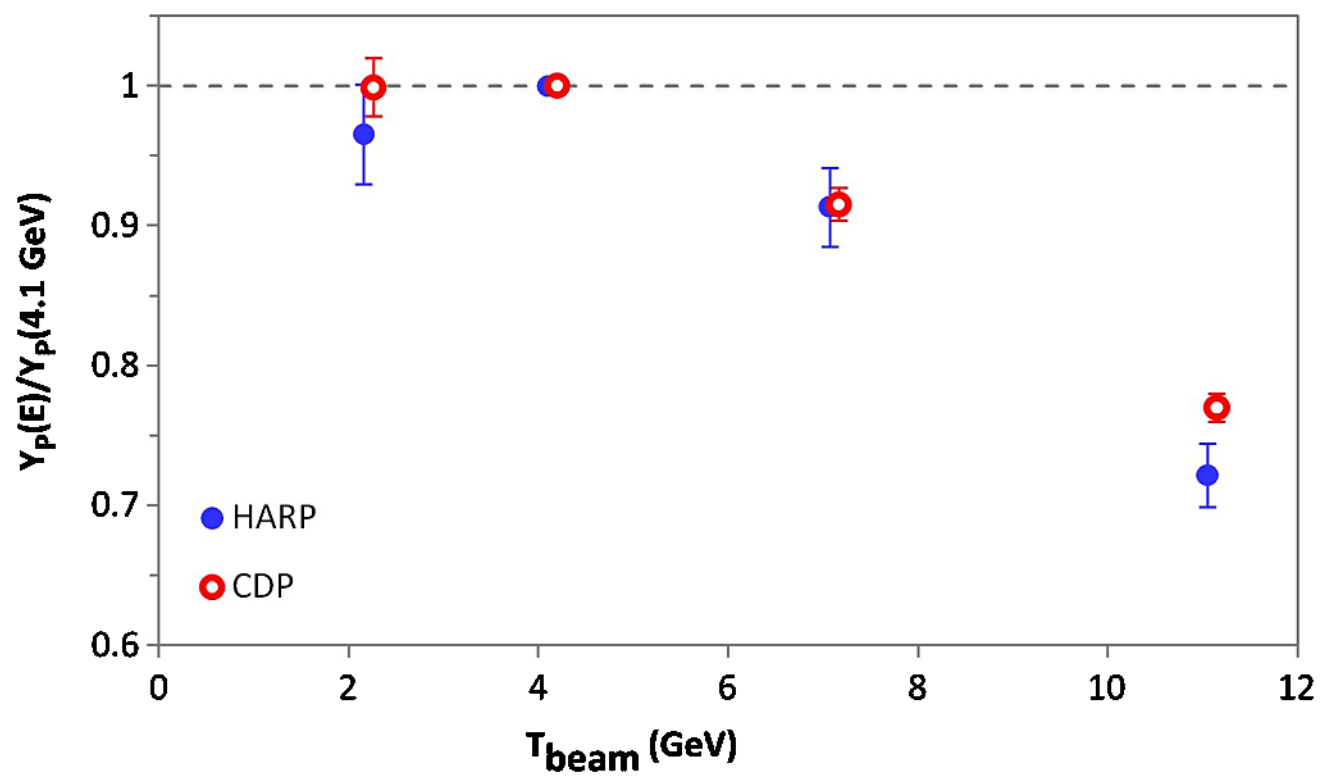

FIG. 6. Relative yield (acceptance-weighted cross section, integrated over the respective phase space areas analyzed by the two HARP groups, divided by beam energy, summed over $\pi^+$ and $\pi^-$ final states, and normalized to the $T_{\rm beam} = 4.1$ GeV value) using the HARP and HARP-CDP results. The points for HARP and CDP have been displaced, respectively, left and right by 50 MeV for clarity of presentation.

$A\sigma^-)/T_{\rm beam}$, where $A\sigma^+(A\sigma^-)$ is the integrated cross section for $\pi^+$ ($\pi^-$) production. This quantity, normalized to the value at 4.1 GeV, is shown in Fig. 6. The single error bar shown here represents the total uncorrelated error: statistical error added in quadrature with the uncorrelated component of the systematic error. The overall yield is maximum at 4.1 GeV, and is within 5% (10%) of the maximum at 2.2 GeV (7.1 GeV). The yield is down by about 25% at 11.1 GeV. The results are essentially the same whether we use the cross section data published by the main HARP group or those published by the HARP-CDP group.

## IV. ESTIMATED CORRECTION FOR PHASE SPACE NOT COVERED BY THE HARP DATA

The results presented by the main HARP group cover momenta $p_\pi > 100$ MeV$/c$. Below this limit, the acceptance of the NF/MC front-end is negligible (see Fig. 3), and the lack of data in this region is not an issue. However, the CDP group presents results with a higher minimum $p_\pi$ of about 150 MeV$/c$ ($p_T < 100$–160 MeV$/c$, depending on $\theta$), and the acceptance is not negligible below their minimum $p_T$. Thus, in the analysis which follows, in which we estimate the correction for the phase space not covered by the HARP data, we consider only the results coming from the main HARP group. In contrast to the dropoff in acceptance at low momentum, the acceptance remains high down to $\theta = 0$. Thus, the lack of data for $\theta < 350$ mrad (20°) cannot be ignored, particularly since the fraction of the total cross section at a small angle is likely to be energy dependent.

Figure 7 shows the acceptance-weighted single differential cross sections divided by beam energy, $A(d\sigma/d\theta)/T_{\rm beam}$, integrated over $p$, using the data from the main HARP group. The error bars represent the total uncorrelated errors [11]. The cross sections at different beam energies have similar shapes. All have a maximum in the $550 < \theta < 750$ mrad bin, and begin to drop towards a smaller angle. This suggests that the fraction of the total cross section below 350 mrad is not too large, and that this fraction is not a strong function of energy. However, a careful examination of the data in Fig. 7 reveals that the slope between the two smallest angle bins is largest for the smallest beam energy, and becomes progressively smaller with increasing beam energy. This suggests that a larger fraction of the total cross section is in the forward region at higher beam energy.

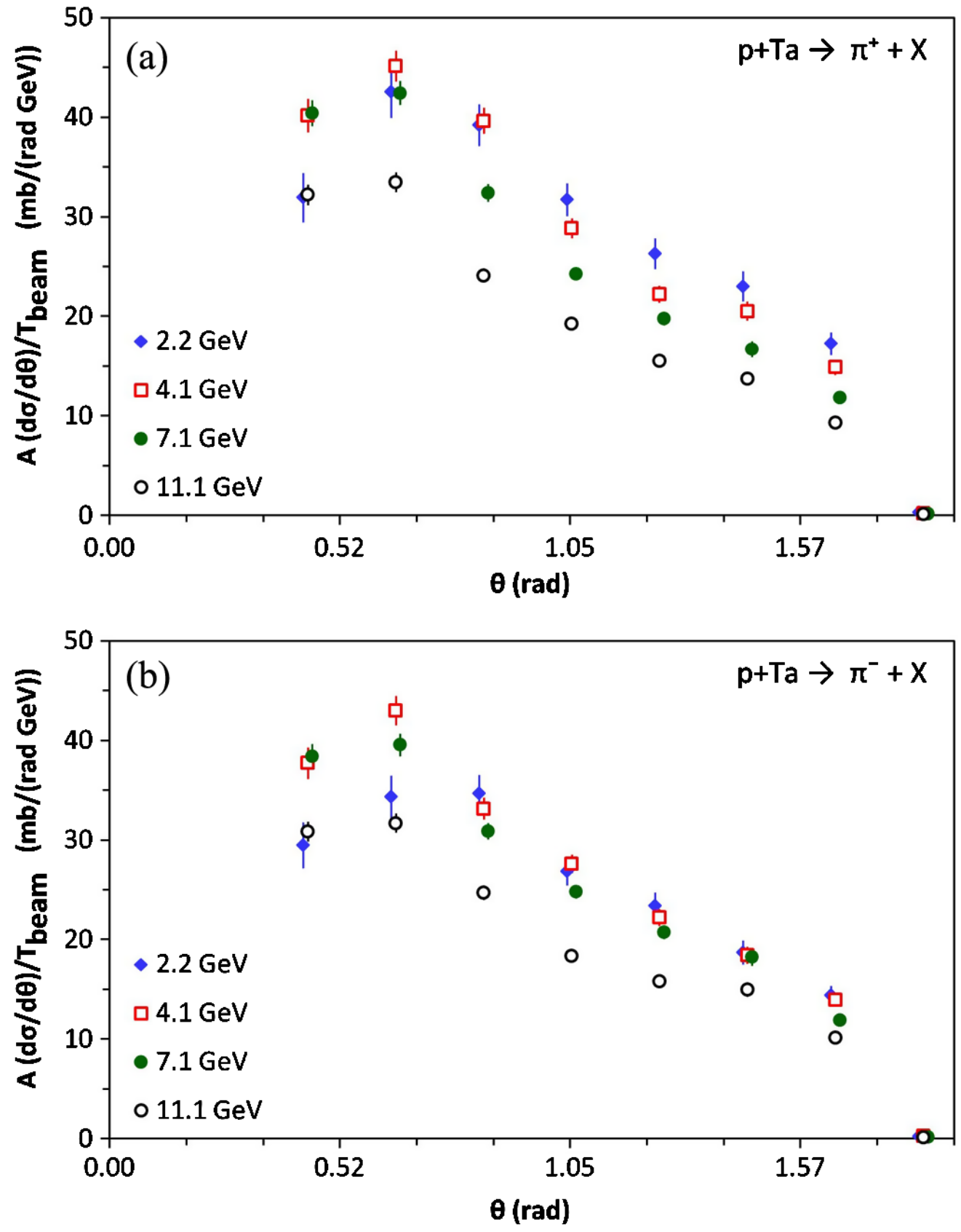

FIG. 7. Acceptance-weighted differential cross section, integrated over momentum, divided by beam energy, using the HARP cross section data for $\pi^+$. The points for $T_{\rm beam} =$ 2.2 GeV and 7.1 GeV have been displaced by $-10$ mrad and $+10$ mrad, respectively, for clarity of presentation. Acceptance-weighted differential cross section, integrated over momentum, divided by beam energy, using the HARP cross section data for $\pi^-$. The points for $T_{\rm beam} = 2.2$ GeV and 7.1 GeV have been displaced by $-10$ mrad and $+10$ mrad, respectively, for clarity of presentation.

To get an indication of the possible magnitude of the energy dependence of the correction for the missing cross section data, we do a simple quadratic extrapolation of the data, constrained to pass through the two lowest angle measurements and $d\sigma/d\theta = 0$ for $\theta = 0$ (since the available phase space $\rightarrow 0$ as $\theta \rightarrow 0$). The integral from 0 to 350 mrad of the extrapolated cross sections gives an estimated increment, $A \cdot \Delta\sigma$, to the measured acceptance-weighted cross sections. The results of the extrapolation are shown in Table I. The estimated increment varies from about 12% at 2.2 GeV to about 19% at 11.1 GeV. The errors shown in the second column of Table I are computed only from the error bars on the two lowest angle data points, and do not take account of any uncertainty that results from the use of this *ad hoc* extrapolation method. The third column of Table I shows how this correction for the unmeasured region would affect the quantity we are interested in, which is the ratio of the acceptance-weighted cross section $A \cdot \sigma$ at a given energy to that at the reference energy of 4.1 GeV. The correction $R_\theta$, amounts to only a few percent. Given the *ad hoc* nature of this correction, we take the uncertainty in $R_\theta$ to be equal to its difference from 1, which is larger than the error derived from the errors in the second column.

TABLE I. Estimated incremental acceptance-weighted cross section, $A \cdot \Delta\sigma$, in the unmeasured region $\theta < 350$ mrad, as a fraction of the measured $A \cdot \sigma_{\rm meas}$ for $\theta > 350$ mrad.

| $T_{\rm beam}$ (GeV) | $A \cdot \Delta\sigma / A \cdot \sigma_{\rm meas}$ | $R_\theta = \frac{(1+A\cdot\Delta\sigma/A\cdot\sigma)}{(1+A\cdot\Delta\sigma/A\cdot\sigma)_{4.1}}$ |
|---|---|---|
| 2.2 | $0.121 \pm 0.014$ | $0.966 \pm 0.034$ |
| 4.1 | $0.160 \pm 0.009$ | 1 |
| 7.1 | $0.187 \pm 0.009$ | $1.023 \pm 0.023$ |
| 11.1 | $0.191 \pm 0.008$ | $1.026 \pm 0.026$ |

## V. CORRECTION FOR THICK TARGET EFFECTS

The HARP data discussed here were taken with thin targets of 5% of a nuclear interaction length ($\lambda_I$), and represent the inclusive cross sections for pion production by protons on a Ta nucleus. In the neutrino factory, a thick target of length 1.5–2 interaction lengths ($\lambda_I$) will be used [1]. In a thick target, there can be secondary interactions of the incident proton and of the outgoing hadrons, i.e., a hadronic shower, which are not accounted for in the pure cross sections discussed above. These secondary interactions can reduce the muon yield through pion absorption, or increase it through creation of additional pions. As shown in Ref. [13], the showering in the target will certainly add additional beam-energy dependence not accounted for in the simple cross section data.

The HARP experiment has measured pion production from thick targets, and the main HARP group has recently published results for carbon, tantalum, and lead targets [14]. They have made corrections for the secondary interactions of the outgoing pions, "such that the effective target is transparent for the secondary 'product' pions and one $1\lambda_I$ long for the 'beam particles'." Thus these data cannot be used directly and additional corrections need to be performed to calculate the effect of reinteraction. This would have to be done in any case because the HARP target was different in diameter (30 mm) and length ($1\lambda_I$) from the target found to be optimal for the NF/MC, 10 mm and $2\lambda_I$, respectively [1,8,9]. We have chosen instead in this paper to estimate the thick vs thin target effects using simulations. The HARP data on thick targets may be used in future studies, particularly once both groups have produced consistent thick target results.

For each of the beam energies for which HARP has presented results, we have run MARS15 simulations, using both the default MARS generator and the LAQGSM [15] generator, for all hadron-nucleus interactions in tantalum target of length $0.05\lambda_I$ and $2\lambda_I$, and diameter = 1 cm. The angle and momentum of each particle exiting the target is recorded, and each pion is weighted by the acceptance of the NF/MC front-end channel, which was previously computed and presented above. The result is the yield $Y$ of muons captured by the front-end channel per incident proton. To take account of the major, but uninteresting effect that more of the incident protons interact in the thick than in the thin target, we then compute the yield per *interacting* beam proton, $Y_I = Y/(1 - e^{-L/\lambda_I})$), where $L$ is the target length. Then this quantity is divided by the beam kinetic energy to give the yield, $Y_P = Y_I/T_{\rm beam}$, normalized to constant proton beam power. The ratio of $R_t = Y_P(L/\lambda_I = 2)/Y_P(L/\lambda_I = 0.05)$ is an effective "amplification factor" due to showering in the target, relative to the simple pion production cross section. Finally, the ratio of $R_t$ at a given beam energy to that at the reference energy of 4.1 GeV is the correction factor for thick target effects, which we apply to the results given in Fig. 5 above.

Table II shows the results of these simulations, summed over $\pi^+$ and $\pi^-$ final states. Each entry in this table is the average of the values computed with the default MARS generator and the LAQGSM generator. The correction for the thick target effects, normalized to 4.1 GeV, is shown in the last column. The error on the correction is taken from the Monte Carlo statistical errors added in quadrature with a systematic error, which we estimate to be ± half the difference between the two simulation results. Naturally, the amplification factor grows with beam energy. The combination of this correction factor and the one for the unmeasured region $\theta < 350$ mrad is plotted in Fig. 8.

TABLE II. Beam power normalized yield of captured muons $Y_P$ calculated by MARS15 for thin and thick tantalum targets, and, in the last column, the "amplification factor" due to showering in the target, relative to the simple pion production cross section, normalized to the value at 4.1 GeV. Results are averages over calculation with the two different event generators.

| $T_{\rm beam}$ (GeV) | $Y_P$ $L/\lambda_I = 0.05$ | $Y_P$ $L/\lambda_I = 2$ | $R_t(T_{\rm beam}) = Y_P(2)/Y_P(0.05)$ | $R_t(T_{\rm beam})/R_t(4.1)$ |
|---|---|---|---|---|
| 2.2 | 0.057 | 0.50 | 0.874 | $0.908 \pm 0.029$ |
| 4.1 | 0.056 | 0.054 | 0.963 | 1.000 |
| 7.1 | 0.053 | 0.057 | 1.079 | $1.122 \pm 0.076$ |
| 11.1 | 0.042 | 0.050 | 1.186 | $1.233 \pm 0.062$ |

Figure 9 shows the yields of muons, 50 m downstream of the target in the NF/MC front-end channel at constant beam power, as a function of beam energy, and normalized to the yield at $T_{beam} = 4.1$ GeV. This figure shows the results with and without corrections for the unmeasured region $\theta < 350$ mrad and the effect of hadronic shower development in a thick target. The points labeled "HARP cross section data" are computed directly from the acceptance-weighted cross section data in Fig. 4, by summing over the two pion charges and dividing by the value at $T_{beam} = 4.1$ GeV. The points labeled "Corrected for $\theta_{min}$ and thick target effects" are similarly computed by multiplying the data in Fig. 4 by the correction factors in Tables I and II, and dividing by the value at $T_{beam} = 4.1$ GeV. The net effect of the corrections is to shift the optimal beam energy from ~4 GeV to ~7 GeV. With the corrections, the beam-power normalized yield at 2 GeV goes from a few percent below that at 4 GeV to 15% below. The yield at 7 GeV goes from 9% below to 5% above that at 4 GeV, and the yield at 11 GeV goes from 28% below to 9% below that at 4 GeV.

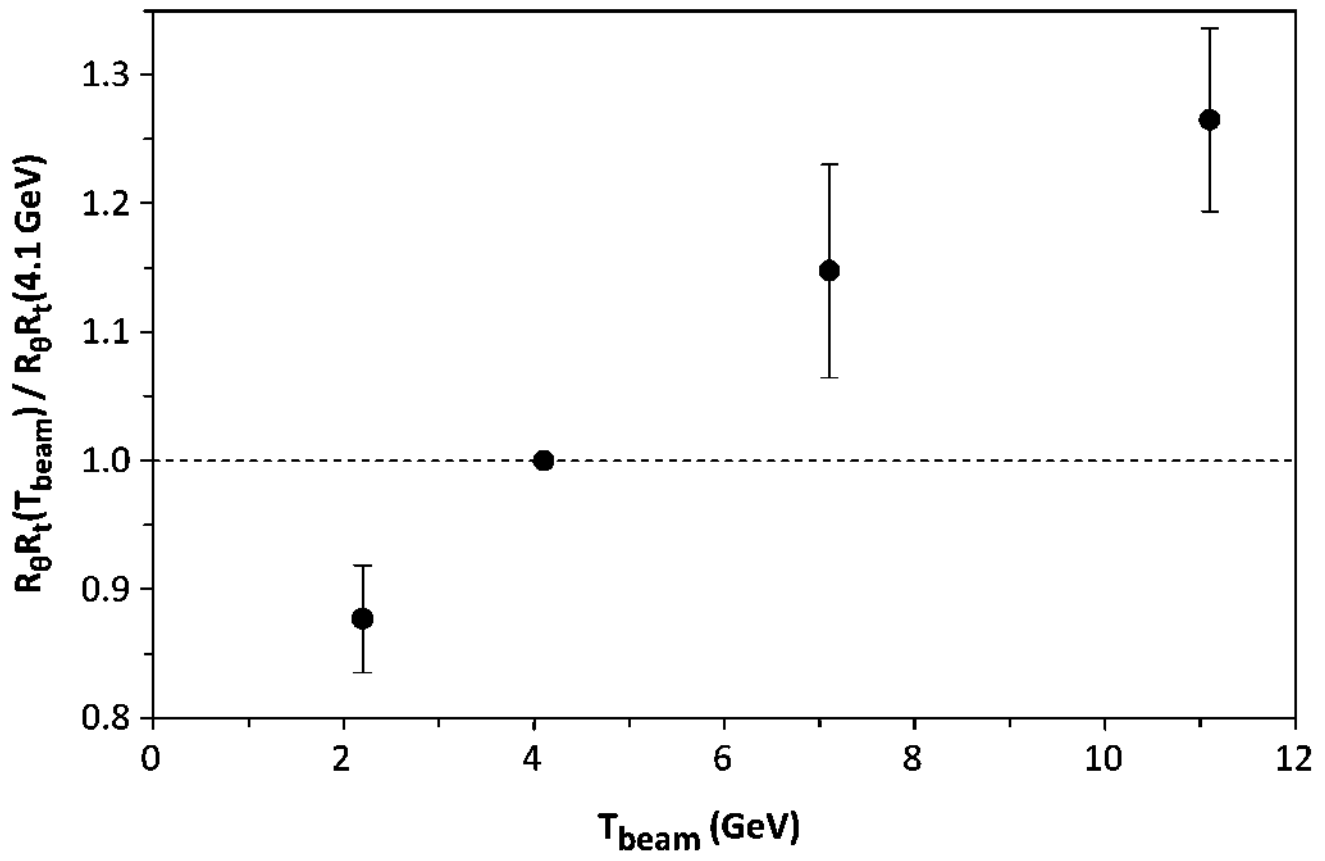

FIG. 8. Combined correction $R_\theta$ for the unmeasured region, $\theta < 350$ mrad, and $R_t$ for the effect of hadronic shower development in a $2\lambda_I$ target, relative to the correction at 4.1 GeV.

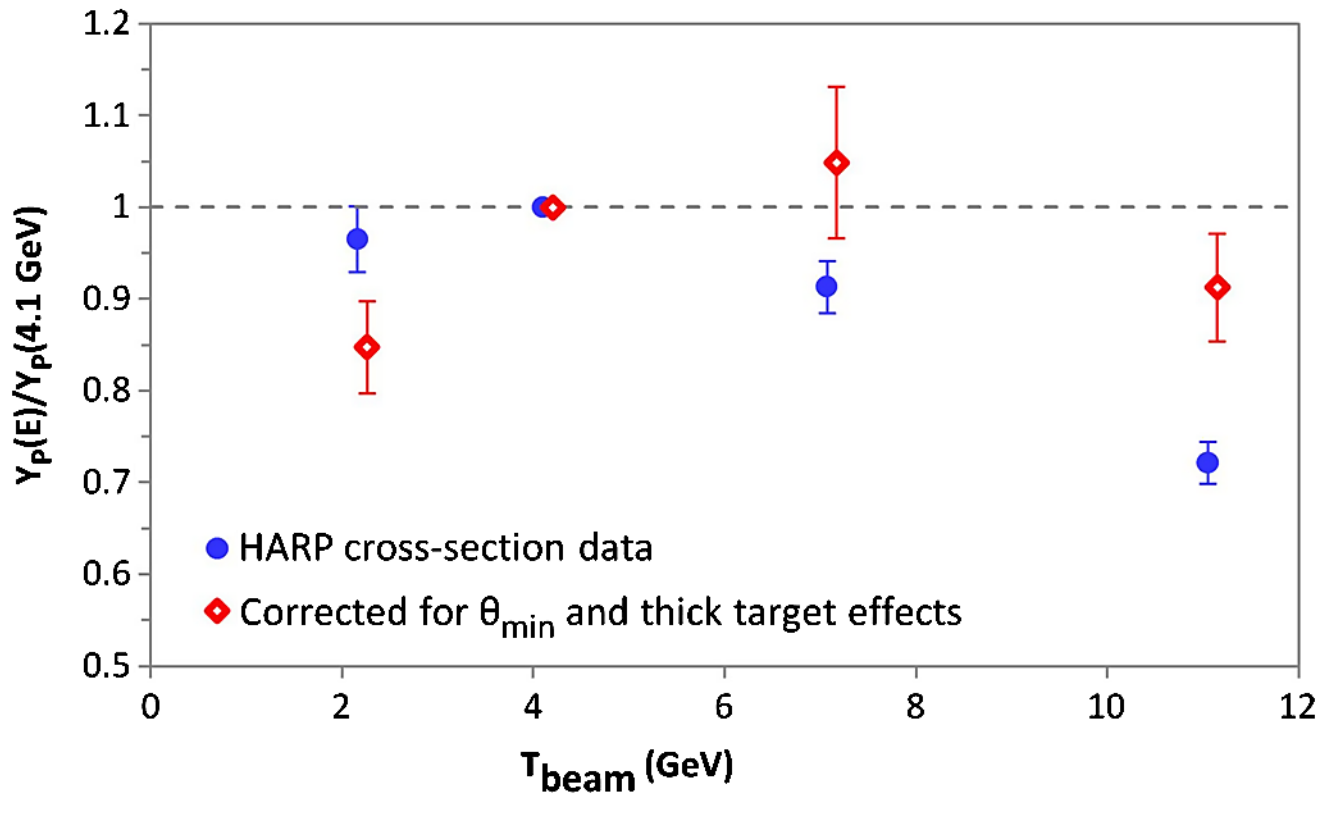

FIG. 9. Beam-power normalized muon yield at the end of the NF/MC front-end channel, relative to that at $T_{beam} = 4.1$ GeV. The results with and without the corrections are displaced +50 and −50 MeV, respectively, for clarity of presentation.

## VI. CONCLUSIONS

The beam energy of the proton driver for a neutrino factory or muon collider is an important design parameter, which can now be chosen based on experimental data from the HARP experiment. In this study we have shown that the measured HARP pion production cross sections, when convoluted with the acceptance of the front-end channel and divided by the beam energy, is maximum for $T_{beam}$ of about 4 GeV, and is within 10% of this maximum for $2 < T_{beam} < 11$ GeV. When this result is corrected for the estimated contribution in the region $\theta < 350$ mrad, which was not measured by HARP, and for the effects of hadronic shower development in a thick target of $\lambda_I = 2$, beam energy giving the largest muon yield, at constant beam power, is about 7 GeV. The dependence of the muon yield on proton beam energy is relatively flat, and any energy between 4 and 11 GeV has a yield that is within 10% of the maximum. In fact, the beam power normalized yield is still 85% of the optimum value for beam energy as low as 2 GeV. These conclusions confirm with a much more complete analysis those reached by HARP themselves in an initial analysis in [3]. They do not depend significantly on whether the results of the main HARP group are used or those from the HARP-CDP group.

One can, therefore, conclude that, from the point of view of muon production and capture, any beam energy in the 4–11 GeV range represents a good choice for the proton driver for a NF/MC. This provides significant latitude in the design of high-power proton sources, which can consider many other optimization parameters than beam energy, such as the ability to concentrate the power in a few bunches or radiation damage issues, without compromising their utility for a neutrino factory or muon collider.

## ACKNOWLEDGMENTS

We would like to thank Roland Garoby for suggesting this analysis to us, and Igor Boyko, Jaap Panman, Gersende Prior, and Jörg Wotschalk for useful discussions that helped us understand how to use the HARP data. This work was supported by Fermi Research Alliance, LLC under Contract No. DE-AC02-07CH11359 with the U.S. Department of Energy.